\documentclass{ws-mpla}
\usepackage[super]{cite}
\usepackage[breaklinks]{hyperref}  
\hypersetup{colorlinks,urlcolor=black,citecolor=black,linkcolor=black,filecolor=black}
\usepackage{breakurl}

\usepackage[utf8]{inputenc}
\usepackage{graphicx}
\usepackage{physics}
\usepackage{amsmath}
\usepackage{comment}
\usepackage{amssymb}

\usepackage{array}
\newcolumntype{M}[1]{>{\raggedright\arraybackslash}m{#1}}
\usepackage{xcolor}

\begin{document}
\markboth{I. Ridkokasha}
{Confronting Dual Models}

\title{CONFRONTING DUAL MODELS OF THE STRONG INTERACTION}
\author{I. RIDKOKASHA}
\address{Faculty of Physics, Taras Shevchenko National University of Kyiv\\
Kyiv, 03127,Ukraine\\
redvan@knu.ua}

\maketitle

\begin{abstract}
    Studies of the mathematical properties of Regge-pole and dual amplitudes are important both for their applications in high energy phenomenology and in their generalizations to strings, superstrings, branes, and other theoretical developments. In the present paper, we investigate the similarities and differences between two classes of dual amplitudes: one with Mandelstam analyticity (DAMA) and another one with logarithmic trajectories (Dual-log). By using quantum (q-) deformations, new features of Dual-log amplitude are unveiled, in particular those concerning its asymptotic behavior and the spectrum of resonances. The two classes of dual amplitudes are compared in various kinematic regions: at fixed transferred momenta asymptotic, fixed angle asymptotic, and in the resonance region. 
\end{abstract}
\begin{quote}
{\fontfamily{cmtt}\selectfont\small
 Electronic version of an article published as Modern Physics Letters A (MPLA) Volume No. 36, Issue No. 05, Article No. 2150031, Year 2021. DOI: \href{https:///dx.doi.org/10.1142/S0217732321500310}{10.1142/S0217732321500310} © copyright World Scientific Publishing Company \href{https:///www.worldscinet.com/mpla}{https:///www.worldscinet.com/mpla}
}
   
\end{quote}
\section{Introduction}
Dual models of the strong interaction, invented by the end of the '60s, opened new trends in high-energy physics. It was initiated using the amazingly simple and attractive Euler Beta function as the scattering amplitude \cite{veneziano68} to unify direct-channel resonances and high-energy $t$-channel Regge-pole description of hadronic reactions.
It evolved into the string theory soon. However, the popularity of dual models significantly decreased (when QCD became the main theory of the strong interactions) for nearly a decade, until its rebirth in the superstring theory\cite{green87}. In this transition, the dual resonance framework was rescaled to describe quantum gravity rather than hadrons\cite{rickles2016brief}. Now string theories have evolved further, into the M-theory, where duality plays a significant role\cite{becker2006string,kaku2012strings,berman2015duality}.

In the present paper, two generalizations of the Veneziano model are studied. Namely, dual amplitude with Mandelstam analyticity (DAMA) and dual amplitude with logarithmic trajectories (Dual-log).

Since then, dual models laid the foundation of the string theory; their further exploration continues and is still relevant\cite{fiore1999fixed,brisudova2000effective,burakovsky1999string}. Dual models are still used in the strong interaction theory for the reaction analysis \cite{shi2015double,szczepaniak2014application}, the confinement problem \cite{biro2018entropy}, and, especially, the pomeron scattering \cite{fiore2016resonance,jenkovszky2011dual,fiore2018exclusive}. One more wondering consequence of dual models is $p$-adic string theory, which presents an unexpected usage of prime numbers in theoretical physics\cite{volovich1987p,freund1987adelic,frampton2020particle,gubser2019spin}.

These works are using both general and specific properties of dual amplitudes, hence results in the present paper may be useful for future discoveries in the areas listed above. Additionally, one may notice that a q-analog in Dual-log amplitude discussed below and q-deformations \cite{hartwig2006deformations,jenkovszky1999quantum,gavrilik2016nonstandard} are similar in their mathematical nature. Therefore, ideas, developed in this work, could be helpful in those questions too.

A summary of similarities and differences between DAMA and Dual-log models could be found in the table \ref{tab1}.

\section{Duality}
Duality imposes strong constraints on the scattering amplitude, namely, resonances in the direct channel sum up in such a way as to produce Regge asymptotic behavior, the number of poles in the $s$- and $t$-planes being infinite. For such an amplitude an integral representation \begin{equation} \label{ansatz} A(s,t)=\int_0^1dxx^{-\alpha(s)-1}f(s,t,x) \end{equation} can be written, where $f(s,t,x)$ is a regular function at $x=0$. Here $s$ and $t$ are the Mandelstam variables.

By expanding $f(s,t,x)$ in a power series around $x=0,$ and integrating this series term by term, one gets 
$$A(s,t)=\int_0^1dxx^{-\alpha(s)-1}\sum_{k=0}^{\infty}a_k(t,s)x^k=\sum_{k=0}^{\infty}\frac{a_k(t,s)}{k-\alpha(s)}. $$
To make this expression crossing-symmetric, one sets $f(s,t,x)=(1-x)^{-\alpha(t)-1}\cdot\\ \cdot g(s,t,x),$ where $g(s,t,x)$ is regular at $x=0$, $x=1$ and symmetric under the interchange $(x,s)\Leftrightarrow(1-x,t)$. By inserting $f(s,t,x)$ in \eqref{ansatz}, one gets the dual amplitude 
$$A(s,t)=\int_0^1dxx^{-\alpha(s)-1}(1-x)^{-\alpha(t)-1}g(s,t,x),$$
or, by setting $g(s,t,x)=1$, its simple version, called the Veneziano amplitude
\begin{align} \label{Veneziano}
A(s,t)&=\int_0^1dxx^{-\alpha(s)-1}(1-x)^{-\alpha(t)-1}=\nonumber\\
&=B(-\alpha(s),-\alpha(t))=
\frac{\Gamma(-\alpha(s))\Gamma(-\alpha(t))}{\Gamma(-\alpha(s)-\alpha(t))}.
\end{align}
The Veneziano model holds only for real, linear trajectories (strings). It opened many new directions in high-energy physics, including strings and superstrings. However, the scattering amplitude \eqref{Veneziano} itself was abandoned and partially forgotten. Paradoxically, the still unsolved problem of the interacting strings was deviated from scattering amplitude \eqref{Veneziano} -- ultimately the (simplified) expression of interacting strings, based on it (for a review of dual and string models see, e.g., Refs.~\refcite{mandelstam73,mandelstam74}).

Possible non-trivial solutions are presented below.

\section{Dual Amplitude with Mandelstam Analyticity (DAMA)}
The $(st)$ term of a dual, crossing-symmetric scattering amplitude with Mandelstam analyticity has the form 
\begin{equation} \label{DAMA}  A(s,t)=\int_0^1
dx\left(\frac{x}{g}\right)^{-\alpha(s')-1}\left(\frac{1-x}{g}\right)^{-\alpha(t')-1},
\end{equation}
where $s'=s(1-x)$, $t'=tx$ ($s$, $t$ are the Mandelstam variables) and $g$ is a parameter ($g>1$). The functions $\alpha(\xi'),\ \xi=s,t,$ called homotopies \cite{bugrij73}, map the physical trajectories $\alpha(\xi,0)=\alpha(\xi)$ onto linear functions $\alpha(\xi,1)=\alpha_0+\alpha'\xi$. (NB: $\xi$ is used to denote $s$ or $t$ to stress the crossing symmetry of the theory.)

The incorporation of complex nonlinear trajectories, impossible in the Veneziano model, becomes mandatory due to the appearance of the integration variable $x$ in the exponents of the integrand in \eqref{DAMA}. Whereas, Mandelstam analyticity (nonvanishing double spectral function, see below) results from the interplay of the integration variable $x$ with the Mandelstam variables, given by the homotopies.

The analytic properties and the asymptotic behavior of the amplitude impose strong bounds on the form of the trajectories, as will be shown below.
\subsection{Asymptotic behavior: polynomial boundedness, Regge, and fixed angle behavior} \label{s31}
Amplitude \eqref{DAMA} is polynomially bounded provided the real part of the trajectories is bounded by
$$\ln \xi \ {\rm Re}(\alpha(\xi))\leq\pi \ {\rm Im}(\alpha(\xi)) \ \  ({\rm here}\ \xi=s\ {\rm or}\  t)$$
or, equivalently, 
$$ \left|\frac{\alpha(\xi)}{\sqrt \xi \ln\xi}\right|_{\xi\rightarrow \infty} \leq \text{const.},\ \ (\xi=s,t).$$

With the above boundary condition, the amplitude for $(s\rightarrow \infty, t = \text{const.})$ is Regge behaved: 
\begin{equation} \label{Regge}
A(s,t)\simeq g^2(-gs)^{\alpha(t)}\left[G(t)+\frac{\ln(-gs)}{s}G_1(t)+
\frac{1}{s}G_2(t)\right], 
\end{equation}
where
$$G(t)=\int_0^{\infty}dz g^{\alpha(-z)}z^{-\alpha(t)-1},$$
$$G_1(t)=t\alpha'(t)\int_0^{\infty}dz
g^{\alpha(-z)}z^{-\alpha(t)},$$
$$G_2(t)=-\left[t\alpha'(t)\int_0^{\infty}dz
g^{\alpha(-z)}z^{-\alpha(t)}\ln z+\int_0^{\infty}dz
g^{\alpha(-z)}z^{-\alpha(t)}(\alpha(-z)+1)\right].$$

Contrary to the Veneziano model \eqref{Veneziano}, associated with linear trajectories (hadronic strings) only, the dual analytic model requires the trajectories to be nonlinear, with a limited real part (limited number of resonances). At the same time, the dual analytic model \eqref{DAMA} reduces to the string model \eqref{Veneziano} in the limit of linear trajectories, or the so-called narrow resonance approximation.

For $(s,|t|\rightarrow\infty,\ s/t=\text{const.}),$ amplitude \eqref{DAMA} scales as
\begin{equation} \label{Dscale} A(s,t)\simeq(st)^{-\gamma\ln(2g)/2}=
e^{-\gamma\ln(2g)\ln(st)/2}, 
\end{equation}
where $\gamma$ is a parameter, iff 
$\alpha(\xi)\sim ln|\xi|,\ \
|\xi|\rightarrow \infty, \ \ 
\xi=s,t.$ 

As it follows, DAMA's asymptotic behavior fundamentally differs for small $t=\text{const.}$ and $t\rightarrow \infty$. In the first kinematic region, the amplitude is Regge behaved; in the latter, it is scaling. The continuous transition between ``soft'' and ``hard'' regimes is an important feature of DAMA. Brief insight about the derivation of the wide-angle asymptotic could be found in the \ref{app1} (for more details see Refs.~\refcite{jenk79,jenk80,fiore1999fixed}).
\subsection{Analytic properties}
The above derivations concerned asymptotic formulae. However, DAMA has all the necessary analytic properties in the low-energy region, i.e., for finite values of its variables. Namely, the trajectory  
\begin{equation} \label{logtr}
\alpha(\xi)=\alpha_0-\alpha_1\ln(1-\alpha_2 \xi)
\end{equation}
is linear for small $|\xi|$ and is logarithmic for large $|\xi|,\ \ \alpha_2|\xi|\gg 1$. Hence, the scattering amplitude \eqref{DAMA} in this ``soft'' limit reproduces the ``stringy'' behavior of the Veneziano model \eqref{Veneziano}: nearly the linear spectrum of resonances and an exponential forward cone, with a dimensional parameter related to the slope of the trajectory. In the ``hard limit'', when both $s$ and $|t|$ are large, their ratio being finite $s/t=\text{const.},$ the amplitude corresponds to the scattering of point-line constituents, obeying quark counting rules.

The low-energy behavior of the amplitude is defined by the analyticity of the $S-$matrix theory. The amplitude has resonance poles corresponding to finite-mass resonances. A typical pole term \eqref{DAMA} has the form
\begin{equation}\label{pole}
g^{k+1}\sum_{l=0}^k[-s\alpha'(s)]^l\frac{C_{k-l}(t)}{[k-\alpha(s)]^{l+1}},
\end{equation}
where 
\begin{equation*}
C_l(t)=\frac{1}{l!}\frac{d^l}{dx_k^l}\left[\left(  \frac{1-x_k}{g} \right)^{-\alpha(tx_k)-1}\right]_{x_k=0}.
\end{equation*}
It follows that $C_l(t)$ is a polynomial of the degree $l$. Formula \eqref{pole} shows that DAMA \eqref{DAMA} does not contain ``ancestors'' (arbitrary heavy resonances for any spin, thus violating unitarity) and that poles of order $(l+1)$ emerge on the daughter levels.

The resonances are reggeized, as was shown by a partial-wave analysis of DAMA \eqref{DAMA}. For trajectories satisfying the asymptotic bound
\begin{equation} \label{bound}
O(\ln^{1+\epsilon}\xi)\leq|\alpha(\xi)|\leq O(\xi^{1/2}),
\end{equation} 
all singularities in the $j$-plane are poles lying on the leading and daughter trajectories and at nonsense points with the wrong signature.

Finite masses can be introduced in the theory through the threshold singularities in the trajectories. Their form is constrained by that of the amplitude (for more details see Ref.~\refcite{bugrij73}). The imaginary part of the trajectory (and of the amplitude) is provided by the light threshold. For example, the two-pion (or a quark-antiquark) threshold can be introduced this way 
\begin{equation} \label{threshold}
\alpha(\xi)=\alpha_0-\alpha_1\sqrt{\xi_0-\xi}, 
\end{equation} 
where the signs are fixed uniquely by the requirement that the imaginary part is positive. Here $\xi_0$ is the threshold value, e.g., $4m_{\pi}^2$ for a bosonic trajectory or $4m_q^2$ for a reggeized gluon ($\xi=s$ or $t$, as before). For infinite\footnote{$m_q\rightarrow\infty$ with the ratio $\alpha'/m_q$ remaining constant} heavy-mass quarks (e.g., those lying in a deep potential well -- the bag) trajectory \eqref{threshold} tends to a linear one, i.e., one reproduces the string mode.

The analytic structure of the amplitude \eqref{DAMA} is rich: apart from direct channel poles, corresponding to Breit-Wigner resonances, and cuts, corresponding to physical thresholds, it also contains a nonvanishing double spectral function $\rho(s,t)$ (double discontinuity) with the boundary
$$(s-s_0)(t-s_0)=s^2$$
required by unitarity. Details can be found in Ref.~\refcite{bugrij73}.
\section{Dual Amplitude with Logarithmic Trajectories (Dual-log)}
Another class of dual models beyond the narrow resonance approximation was suggested by D.D. Coon, S. Yu, and M. Baker \cite{coon72}. It was developed further and compared with data in Refs.~\refcite{arik74,arik75} by M. Arik. This class of dual models is intimately connected with logarithmic trajectories.

Now let us consider one concrete example of dual amplitude introduced by M. Arik and D.D. Coon \cite{arik74}.
In the original work, it was presented in such a form:
\begin{equation}\label{dual-log}
    M = -q^{\alpha_s \alpha_t}\frac{G(q)G(q/\sigma \tau)}{G(q/\sigma)G(q/\tau)},
\end{equation}
where
\begin{equation*}
    \alpha_s = \frac{\ln \sigma}{\ln q}, \; \alpha_t = \frac{\ln \tau}{\ln q}, \; |q|<1,
\end{equation*}
\begin{equation*}
    G(z) = \prod_{l=0}^{\infty}(1-q^lz).
\end{equation*}
Amplitude is symmetric between $\sigma$ and $\tau$. If $\sigma = b - as$ and $\tau = b - at$, this implies crossing symmetry between $s$ and $t$. On the other hand, for processes that are not crossing symmetric, $\sigma$ and $\tau$ could have different linear dependencies on $s$ and $t$. Below this amplitude will be considered in cross-symmetrical form with $\sigma = b - as$, $\tau = b - at$.

For investigation of this model, it will be helpful to become acquainted with the q-analog, especially the q-Pochhammer symbol.
\subsection{q-analog}
In mathematics, a q-analog of a theorem, identity, or expression is a generalization involving a new parameter q that returns the original theorem, identity, or expression in the limit as $q\rightarrow 1$. (More details could be found in Refs.~\refcite{andrews99,gasper04})

Classical q-theory begins with the q-analogs of the nonnegative integers. The equality
\begin{equation*}
\lim_{q\rightarrow 1}\frac{1-q^n}{1-q}=n
\end{equation*}
suggests that one defines the q-analog of n, also known as the q-bracket or q-number of n, to be
\begin{equation*}
    [n]_q=\frac{1-q^n}{1-q} = 1 + q + q^2 + \ldots + q^{n - 1}.
\end{equation*}

However, the most useful q-analog for this work is a q-Pochhammer symbol -- a q-analog of the Pochhammer symbol. Its definition is 
\begin{equation*}
    (a;q)_n = \prod_{k=0}^{n-1} (1-aq^k)=(1-a)(1-aq)(1-aq^2)\cdots(1-aq^{n-1}).
\end{equation*}
Unlike the ordinary Pochhammer symbol, the q-Pochhammer symbol can be extended to an infinite product:
\begin{equation*}
    (a;q)_\infty = \prod_{k=0}^{\infty} (1-aq^k).
\end{equation*}
Relation between the finite and the infinite q-Pochhammer symbols:
\begin{equation}\label{q-Poch}
    (aq^n;q)_\infty= \frac{(a;q)_\infty} {(a;q)_n }.
\end{equation}

Another important identity is the so-called q-binomial theorem, proof of which could be found in Ref.~\refcite{andrews99}.
If $|x|<1$ and $|q|<1$, then
\begin{equation} \label{origin_q-bin}
    \frac{(ax;q)_\infty}{(x;q)_\infty} = \sum_{n=0}^\infty \frac{(a;q)_n}{(q;q)_n} x^n.
\end{equation}
Multiplying both parts by a factor 
$\frac{(q;q)_\infty}{(a;q)_\infty}$
and using \eqref{q-Poch} it takes form
\begin{equation} \label{q-bin}
    \frac{(ax;q)_\infty (q;q)_\infty}{(x;q)_\infty (a;q)_\infty} =\sum_{n=0}^\infty \frac{ (q;q)_\infty/(q;q)_n}{ (a;q)_\infty/(a;q)_n} x^n = \sum_{n=0}^\infty \frac{(q^{n+1};q)_\infty}{(q^na;q)_\infty} x^n.
\end{equation}

Another interesting formula is an approximation for $(x;q)_\infty$ when $x\rightarrow 0$\footnote{The same identity could be obtained after substitution $a = 0$ to the \eqref{origin_q-bin} and neglecting higher-order terms}:
\begin{equation} \label{q-Poch_approx}
  (x;q)_\infty = (1-x)(1-xq)(1-xq^2)... \simeq 1-x - xq - xq^2  - ... = 1-\frac{x}{1-q}.
\end{equation}

Additionally, an infinite q-Pochhammer symbol could be rewritten into the infinite sum, using exponent properties:
\begin{equation*}
    (q^x;q)_\infty = \exp\left[\sum_{k=0}^\infty \ln{(1-q^{x+k})}\right]. 
\end{equation*}
Using logarithm series expansion for $x>0$ this identity takes such a form:
\begin{equation*}
    (q^x;q)_\infty = \exp\left[-\sum_{k=0}^\infty \sum_{m=1}^\infty \frac{(q^{x+k})^m}{m}\right].
\end{equation*}
By changing a summation order and using an infinite geometric progression sum formula
\begin{equation*}
    \exp\left[-\sum_{m=1}^\infty \frac{q^{xm}}{m} \sum_{k=0}^\infty (q^m)^k\right]
    =  \exp\left[-\sum_{m=1}^\infty \frac{q^{xm}}{m(1-q^m)}\right].
\end{equation*}
Therefore, for $x > 0$
\begin{equation} \label{exponent}
    (q^x;q)_\infty
    =  \exp\left[-\sum_{m=1}^\infty \frac{q^{xm}}{m(1-q^m)}\right].
\end{equation}
\subsection{Important identities for Dual-log amplitude}
Equation \eqref{dual-log} could be rewritten using q-analog notation in these forms:
\begin{equation} \label{q-dual-log}
    M = -\tau^{\ln \sigma/\ln q}\frac{(q;q)_\infty (q/\sigma \tau;q)_\infty}{(q/\sigma;q)_\infty (q/\tau;q)_\infty} = -q^{\alpha_s \alpha_t}\frac{(q;q)_\infty (q^{1-\alpha_s-\alpha_t};q)_\infty}{(q^{1-\alpha_s};q)_\infty (q^{1-\alpha_t};q)_\infty}.
\end{equation}
Amplitude could be transformed into infinite series, using the q-binomial theorem, namely, \eqref{q-bin} identity. First, we rewrite \eqref{q-dual-log} equation into this form:
\begin{equation*}
    M =  -\tau^{\ln \sigma/\ln q}\frac{(1-1/\sigma)(1-1/\tau)}{1-1/\sigma \tau}\frac{(q;q)_\infty (1/\sigma \tau;q)_\infty}{(1/\sigma;q)_\infty (1/\tau;q)_\infty}.
\end{equation*}
Now, using \eqref{q-bin} equation, where $a = 1/\sigma$ and $x = 1/\tau$, 
\begin{equation} \label{t-series}
    M = -\tau^{\ln \sigma/\ln q}\frac{(1-1/\sigma)(1-1/\tau)}{1-1/\sigma \tau}\sum_{n=0}^\infty \frac{(q^{n+1};q)_\infty}{(q^n /\sigma;q)_\infty} (1/\tau)^n,
\end{equation}
for $|1/\tau|<1$.
Similarly, from the symmetry of amplitude between $\tau$ and $\sigma$ for $|1/\sigma|<1$
\begin{equation} \label{s-series}
    M = -\sigma^{\ln \tau/\ln q}\frac{(1-1/\sigma)(1-1/\tau)}{1-1/\sigma \tau}\sum_{n=0}^\infty \frac{(q^{n+1};q)_\infty}{(q^n /\tau;q)_\infty} (1/\sigma)^n.
\end{equation}

Moreover, an amplitude could be presented as an exponent of the infinite sum, using equation \eqref{exponent}. For $\alpha_s<1$, $\alpha_t<1$, and $\alpha_s+\alpha_t<1$, an amplitude \eqref{q-dual-log} transforms into
\begin{align*}
    M &= -q^{\alpha_s \alpha_t}\exp\left[-\sum_{m=1}^\infty \frac{q^{m}}{m(1-q^m)}(1+q^{-(\alpha_s+\alpha_t)m}-q^{-\alpha_s m}-q^{-\alpha_t m})\right] =\\
   &=-\sigma^{\ln \tau/\ln q}\exp\left[-\sum_{m=1}^\infty \frac{q^{m}}{m(1-q^m)}(1+(1/\sigma\tau)^{m}-(1/\sigma)^{m}-(1/\tau)^{m})\right].
\end{align*}
Similarly, using equations \eqref{exponent} and \eqref{t-series} for $\alpha_s < 0$ could be obtained
\begin{align*}
    M &= -\tau^{\ln \sigma/\ln q}\frac{(1-1/\sigma)(1-1/\tau)}{1-1/\sigma \tau} \\ 
    &\qquad\sum_{n=0}^\infty (1/\tau)^n \exp\left[-\sum_{m=1}^\infty \frac{q^{mn}}{m(1-q^m)}(q^m-q^{-m\alpha_s})\right] = \\
    &=-\tau^{\ln \sigma/\ln q}\frac{(1-1/\sigma)(1-1/\tau)}{1-1/\sigma \tau}\\
    &\qquad\sum_{n=0}^\infty (1/\tau)^n \exp\left[-\sum_{m=1}^\infty \frac{q^{mn}}{m(1-q^m)}(q^m-(1/\sigma)^{m})\right].
\end{align*}
Correspondingly, from expressions \eqref{exponent} and \eqref{s-series} for $\alpha_t < 0$
\begin{align*}
    M &=  -\sigma^{\ln \tau/\ln q}\frac{(1-1/\sigma)(1-1/\tau)}{1-1/\sigma \tau}\\
    &\qquad\sum_{n=0}^\infty (1/\sigma)^n \exp\left[-\sum_{m=1}^\infty \frac{q^{mn}}{m(1-q^m)}(q^m-q^{-m\alpha_t})\right]= \\
    &=-\sigma^{\ln \tau/\ln q}\frac{(1-1/\sigma)(1-1/\tau)}{1-1/\sigma \tau}\\
    &\qquad\sum_{n=0}^\infty (1/\sigma)^n \exp\left[-\sum_{m=1}^\infty \frac{q^{mn}}{m(1-q^m)}(q^m-(1/\tau)^{m})\right] .
\end{align*}
\subsection{Analytic properties}\label{s43}
Amplitude \eqref{q-dual-log} coincides (up to a factor) with the so-called q-Beta function. To check it we calculate the limit $q\rightarrow 1$ (as all $q$-analogs):
\begin{align*}
\lim_{q\rightarrow 1}M&=
\lim_{q\rightarrow 1}-q^{\alpha_s \alpha_t}\frac{(q;q)_\infty (q^{1-\alpha_s-\alpha_t};q)_\infty}{(q^{1-\alpha_s};q)_\infty (q^{1-\alpha_t};q)_\infty}=\\
&=\lim_{q\rightarrow 1}-q^{\alpha_s \alpha_t}\prod_{n=1}^{\infty}\frac{(1-q^n)(1-q^{n-\alpha_s-\alpha_t})}{(1-q^{n-\alpha_s}) (1-q^{n-\alpha_t})}.
\end{align*}
Changing the variables $q=1-u$, one obtains
\begin{align*}
\lim_{q\rightarrow 1}M&=\lim_{u\rightarrow 0}-(1-u)^{\alpha_s \alpha_t}\prod_{n=1}^{\infty}\frac{(1-(1-u)^n)(1-(1-u)^{n-\alpha_s-\alpha_t})}{(1-(1-u)^{n-\alpha_s}) (1-(1-u)^{n-\alpha_t})} =\\
&=-\prod_{n=1}^{\infty}\frac{n(n-\alpha_s-\alpha_t)}{(n-\alpha_s) (n-\alpha_t)}.
\end{align*}
For the beta function, there is a well-known formula
\begin{equation*}
B(x,y)=\frac{x+y}{xy}\prod_{n=1}^{\infty}\frac{n(n+x+y)}{(n+x) (n+y)}.
\end{equation*}
Consequently, 
\begin{equation*}
\lim_{q\rightarrow 1}M = \frac{\alpha_s\alpha_t}{\alpha_s+\alpha_t} B(-\alpha_s;-\alpha_t).
\end{equation*}
This equation shows the relationship between Dual-log amplitude \eqref{dual-log} and Veneziano amplitude \eqref{Veneziano}.

The next interesting property is pole behavior. From the infinite product representation of an amplitude
\begin{equation*}
M = -q^{\alpha_s \alpha_t}\prod_{n=1}^{\infty}\frac{(1-q^n)(1-q^{n-\alpha_s-\alpha_t})}{(1-q^{n-\alpha_s}) (1-q^{n-\alpha_t})},
\end{equation*}
one finds that it has singularities when $(1-q^{n-\alpha_s}) (1-q^{n-\alpha_t})=0$. Hence, all poles correspond to $\alpha_{s,t} = k , k \in \mathbb{N} $. To investigate further, let us consider $\alpha_s \rightarrow k  $. The only term that leads to a discontinuity at this point is $1/(1-q^{k-\alpha_s})$. It could be approximately rewritten in  such a way:
\begin{equation*}
\frac{1}{1-q^{k-\alpha_s}}\simeq \frac{1}{1-(1+(k-\alpha_s)\ln q)} = \frac{1}{(\alpha_s-k)\ln q}.
\end{equation*}
Therefore, for $\alpha_s \rightarrow k$
\begin{align*}
M &= -q^{\alpha_s \alpha_t}\frac{(q;q)_\infty (q^{1-\alpha_s-\alpha_t};q)_\infty}{(q^{1-\alpha_t};q)_\infty} \prod_{n=1}^{\infty}\frac{1}{(1-q^{n-\alpha_s})}\simeq\\
&\simeq-q^{k \alpha_t}\frac{(q;q)_\infty (q^{1-k-\alpha_t};q)_\infty}{(q^{1-\alpha_t};q)_\infty} \frac{1}{(\alpha_s-k)\ln q}\prod_{n=1,n\neq k}^{\infty}\frac{1}{(1-q^{n-k})}.
\end{align*}
Using equation \eqref{q-Poch} and the definition of the q-Pochhammer symbol it simplifies to
\begin{equation}\label{pole_dual-log}
M = -\frac{q^{k\alpha_t}}{(\alpha_s-k)\ln q} \frac{(q^{1-k-\alpha_t};q)_k}{(q^{1-k};q)_{k-1}}.
\end{equation}

Hence, all poles of Dual-log amplitude are of order 1, correspond to $\alpha_{s,t} = k , k \in \mathbb{N}$, and could be calculated using equation \eqref{pole_dual-log}.
\subsection{Asymptotic behavior}

To find an asymptotic ($s\rightarrow\infty,\ t=\text{const.}$) in the Dual-log model, we use \eqref{s-series} formula:
\begin{equation*}
    M = -\sigma^{\ln \tau/\ln q}\frac{(1-1/\sigma)(1-1/\tau)}{1-1/\sigma \tau}\sum_{n=0}^\infty \frac{(q^{n+1};q)_\infty}{(q^n /\tau;q)_\infty} (1/\sigma)^n.
\end{equation*}
Conditions ($s\rightarrow\infty,\ t=\text{const.}$) are equivalent to ($1/\sigma \rightarrow 0$ and $|1/\tau|=\text{const.}$). Therefore, the zero-order approximation looks like
\begin{equation} \label{Regge-dual-log0}
    M_{(0)}^s = -\sigma^{\ln \tau/\ln q}\frac{(1-1/\tau)(q;q)_\infty}{(1 /\tau;q)_\infty}.
\end{equation}
The first-order:
\begin{align}\label{Regge-dual-log1}
    M_{(1)}^s &= -\sigma^{\ln \tau/\ln q} (1-1/\tau) (1-1/\sigma+1/\sigma\tau)\nonumber\\
    &\qquad\left[\frac{(q;q)_\infty}{(1 /\tau;q)_\infty} + \frac{(q^2;q)_\infty}{(q /\tau;q)_\infty} (1/\sigma)\right] = \nonumber\\
    &=-\sigma^{\ln \tau/\ln q} \frac{(1-1/\tau)(q;q)_\infty}{(1 /\tau;q)_\infty} \nonumber\\
    &\qquad\left[1+\frac{1}{\sigma}\left(-1+\frac{1}{\tau}+\frac{(q^2;q)_\infty (1 /\tau;q)_\infty}{(q;q)_\infty (q /\tau;q)_\infty}\right)\right] = \nonumber\\ 
    &=M_{(0)}^s\left[1+\frac{1}{\sigma}\left(-1+\frac{1}{\tau}+\frac{1-1/\tau}{1-q}\right)\right]= \nonumber\\
    &= M_{(0)}^s\left[1+\frac{1}{\sigma}\frac{q}{1-q}\left(1-\frac{1}{\tau}\right)\right].
\end{align}

For $(s,|t|\rightarrow\infty,\ s/t=\text{const.})$ or, similarly, $(\sigma, \tau \rightarrow \infty, \; \sigma/\tau < \text{const.}),$ Dual-log amplitude could be found from previous asymptotic \eqref{Regge-dual-log0}, when $1/\tau$ approaches $0$
\begin{equation}\label{scale-dual-log0}
    M_{(0)}^s = -\sigma^{\ln \tau/\ln q}\frac{(1-1/\tau)(q;q)_\infty}{(1 /\tau;q)_\infty} \simeq -e^{\ln \sigma\ln \tau/\ln q}(q;q)_\infty=M_{(0)}^{st}.
\end{equation}
Using equations \eqref{Regge-dual-log1} and \eqref{q-Poch_approx}
\begin{align} \label{scale-dual-log1}
    M_{(1)}^s &= 
    -\sigma^{\ln \tau/\ln q}\frac{(1-1/\tau)(q;q)_\infty}{(1 /\tau;q)_\infty}\left[1+\frac{1}{\sigma}\frac{q}{1-q}(1-1/\tau)\right] \simeq \nonumber\\ 
    &\simeq-e^{\ln \sigma \ln \tau/\ln q}\frac{(1-1/\tau)(q;q)_\infty}{1-\frac{1/\tau}{1-q}}\left[1+\frac{1}{\sigma}\frac{q }{1-q}\right] \simeq  \nonumber\\
    &\simeq-e^{\ln \sigma \ln \tau/\ln q}(q;q)_\infty\left[1 - 1/\tau +\frac{1/\tau}{1-q}+\frac{1}{\sigma}\frac{q }{1-q}\right] = \nonumber\\
    &=-e^{\ln \sigma \ln \tau/\ln q}(q;q)_\infty\left[1 + \frac{q }{1-q}\left(\frac{1}{\tau}+\frac{1}{\sigma}\right)\right]=M_{(1)}^{st}.
\end{align}
\section{Resonances and Asymptotic Behavior}\label{s5}
In this section, two classes of dual model, DAMA and Dual-log, are discussed in various kinematic regions: fixed transferred momenta ($s\rightarrow\infty,\ t=\text{const.}$), fixed angle ($s,t\rightarrow\infty,\ s/t=\text{const.}$), and resonances ($s\rightarrow s_{res}$).    
\subsection{Fixed transferred momenta asymptotic ($s\rightarrow\infty,\ t=\text{const.}$)}
For comparison, we rewrite Dual-log amplitude approximations \eqref{Regge-dual-log0} and \eqref{Regge-dual-log1} in terms of $s$ and $t$ using equalities $\sigma = b - as$ and $\tau = b - at$:
\begin{align} \label{dual-log_Regge_asym}
M_{(0)}^s &\simeq -(-as)^{\alpha_t}\frac{(1-1/(b-at))(q;q)_\infty}{(1 /(b-at);q)_\infty},\nonumber\\
M_{(1)}^s &= -(-as)^{\alpha_t}(1-b/as)^{\alpha_t}\frac{(1-1/(b-at))(q;q)_\infty}{(1 /(b-at);q)_\infty}\nonumber\\ &\qquad \left[1 - \frac{1}{as}\frac{1}{1-b/as}\frac{q}{1-q}\left(1-\frac1{b-at}\right)\right]
\simeq \nonumber\\
&\simeq-(-as)^{\alpha_t}\frac{(1-1/(b-at))(q;q)_\infty}{(1 /(b-at);q)_\infty}\nonumber\\ 
&\qquad\left[1- \frac{1}{as}\left(  b\alpha_t + \frac{q}{1-q}\left(1-\frac1{b-at}\right)\right)\right].
\end{align}
The reason why the term $q^{\alpha_s\alpha_t}$ was introduced in \eqref{dual-log} is clear from \eqref{dual-log_Regge_asym}. This term is responsible for the Regge asymptotic of the Dual-log amplitude. Additionally, it does not spoil the limit to the Veneziano model when $q\rightarrow 1$.

Compared to \eqref{Regge} in DAMA
\begin{equation*} 
A(s,t)\simeq g^2(-gs)^{\alpha(t)}\left[G(t)+\frac{\ln(-gs)}{s}G_1(t)+
\frac{1}{s}G_2(t)\right], \end{equation*} where
$$G(t)=\int_0^{\infty}dz g^{\alpha(-z)}z^{-\alpha(t)-1},$$
$$G_1(t)=t\alpha'(t)\int_0^{\infty}dz
g^{\alpha(-z)}z^{-\alpha(t)},$$
$$G_2(t)=-\left[t\alpha'(t)\int_0^{\infty}dz
g^{\alpha(-z)}z^{-\alpha(t)}\ln z+\int_0^{\infty}dz
g^{\alpha(-z)}z^{-\alpha(t)}(\alpha(-z)+1)\right].$$

Regge asymptotic behavior restricts zero-order approximation of both amplitudes to $(-s)^{\alpha(t)}$. However, already in the next term, the two models have differences. DAMA has the $\ln(-s)$ term, but Dual-log does not (because Dual-log amplitude could be rewritten in the Taylor series using the q-binomial theorem \eqref{s-series}).
Therefore, the only similarity between these two amplitudes is that they both satisfy the Regge behavior requirement. 

The transformation of Regge asymptotic when $t\rightarrow \infty$ is more significant and is discussed below.
\subsection{Fixed angle asymptotic ($s,t\rightarrow\infty,\ s/t=\text{const.}$)}
Similar to the previous asymptotic, we rewrite Dual-log amplitude approximations \eqref{scale-dual-log0} and \eqref{scale-dual-log1} in terms of $s$ and $t$.

\begin{align} \label{dual-log_fixed_asym}
    M_{(0)} &\simeq -e^{\ln(-as) \ln(-at)/\ln q}(q;q)_\infty, \nonumber\\   
    M_{(1)} &= -e^{(\ln(-as) + \ln(1 - b/as ))(\ln(-at) + \ln(1 - b/at ))/\ln q}(q;q)_\infty\nonumber\\
    &\qquad\left[1 + \frac{q }{1-q}\left(\frac{1}{-at}\frac{1}{1-b/at}+\frac{1}{-as}\frac{1}{1-b/as}\right)\right]  \simeq \nonumber\\
    &\simeq-e^{\ln(-as) \ln(-at)/\ln q}(q;q)_\infty(1 - b/as )^{\ln(-at)}(1 - b/at )^{\ln(-as)}\nonumber\\
    &\qquad\left[1 + \frac{q }{1-q}\left(\frac{1}{-at}+\frac{1}{-as}\right)\right] \simeq \nonumber\\
    &\simeq-e^{\ln(-as) \ln(-at)/\ln q}(q;q)_\infty\nonumber\\
    &\qquad\left[1 - \frac{b}{a\ln q}\left(\frac{\ln(-at)}{s}+\frac{\ln(-as)}{t}\right) -\frac{q }{(1-q)a}\left(\frac{1}{s} + \frac{1}{t}\right)\right].
\end{align}

The Dual-log fixed angle asymptotic is similar to the fixed transferred momenta asymptotic. Even when both $t$ and $s$ approach infinity, amplitude behavior stays Regge $M \sim (-as)^{\ln(-at)/\ln q}=(-at)^{\ln(-as)/\ln q}$. However, DAMA's behavior is far less straightforward, as was mentioned in subsection \ref{s31}. Its Regge behavior is restricted by finite $t$. 
For $(s,|t|\rightarrow\infty,\ s/t=\text{const.}),$ DAMA scales as
\begin{equation}
A(s,t)\simeq(st)^{-\gamma\ln(2g)/2}=
e^{-\gamma\ln(2g)\ln(st)/2}, 
\end{equation}
where $\gamma$ is a parameter.

Although DAMA's wide-angle behavior is not connected directly to that derived from QCD, still the final formula can be compared with those derived from pQCD. What is more important, DAMA realizes a smooth transition from typically ``soft'' (non-perturbative) behavior to the ``hard'' (perturbative) regime. However, this behavior is not inherent in the Dual-log model.
\subsection{Resonances}
In Dual-log all singularities correspond to $\alpha_{s,t} = k , k \in \mathbb{N} $. These singularities are simple poles \eqref{pole_dual-log}:
\begin{equation}
M = -\frac{q^{k\alpha_t}}{(\alpha_s-k)\ln q} \frac{(q^{1-k-\alpha_t};q)_k}{(q^{1-k};q)_{k-1}}.
\end{equation}
In DAMA singularities' behavior is far more complex (for details see Ref.~\refcite{bugrij73}). Nevertheless, DAMA too has poles for $\alpha_{s,t} = k , k \in \mathbb{N}$, but of order $k+1$ \eqref{pole}:

$$g^{k+1}\sum_{l=0}^k[-s\alpha'(s)]^l\frac{C_{k-l}(t)}{[k-\alpha(s)]^{l+1}}.$$

\begin{figure}[h!]
\begin{center}
\begin{minipage}[h]{0.7\linewidth}
\includegraphics[width=1\linewidth]{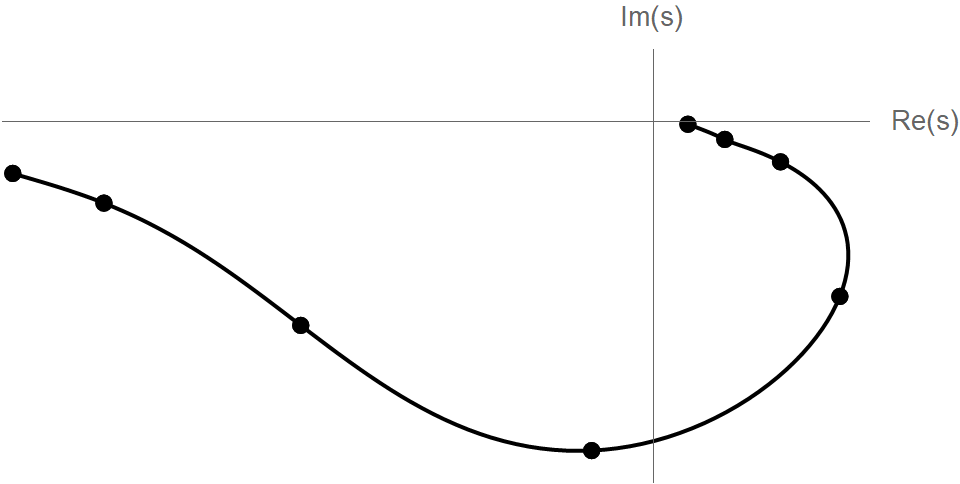}
\caption{DAMA poles in the s plane.}
\label{fig:DAMA}
\end{minipage}
\begin{minipage}[h]{0.7\linewidth}
\includegraphics[width=1\linewidth]{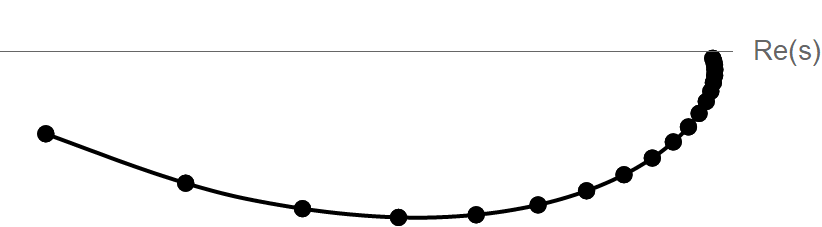}
\caption{Dual-log poles in the s plane.}
\label{fig:Dual}
\end{minipage}
\end{center}
\end{figure}

Therefore, the two models have poles for the same values of trajectories ($\alpha_{s,t} = k , k \in \mathbb{N} $). It is a consequence of their common with the Veneziano amplitude, which has the same poles. However, the positions of poles in the $s$ plane do not coincide because amplitudes have different trajectories (figures \ref{fig:DAMA} and \ref{fig:Dual}). For DAMA, see Ref.~\refcite{bugrij73}; for Dual-log, it is easy to derive. For resonances $\alpha_s=k$, from \eqref{dual-log} $\alpha_s=\frac{\ln (b-as)}{\ln q}$. Hence:
$$s=\frac{b-q^{\alpha_s}}{a}=\frac{b-q^k}{a}.$$

Among the basic physical implications that DAMA and Dual-log models aim to describe, the one is - do resonances melt, transforming into a quark-gluon soup (see Ref.~\refcite{biro2018entropy}). It follows from DAMA that the number of resonances is finite - they melt beyond a certain mass. In Dual-log models, this is the ``ionization point''.

\section{Conclusions}
In this paper, two classes of dual models (Dual-log and DAMA) are studied. Both amplitudes arose from the generalizations of the Veneziano model. 

In the table \ref{tab1} we briefly list DAMA's and Dual-log's properties, their differences, and similarities (for details, see section \ref{s5}). Each cell represents the corresponding amplitude form.
\renewcommand{\arraystretch}{3}
\begin{table}[h]
    \centering
    \tbl{Summary of DAMA's and Dual-log's features}
    {
    \begin{tabular}{|M{10.5em}|m{19em}|m{12.5em}|}
    
		 \hline
         &\textbf{DAMA}&\textbf{Dual-log}\\
         \hline
       Scattering amplitude & $\int_0^1dx\left(\frac{x}{g}\right)^{-\alpha(s')-1}\left(\frac{1-x}{g}\right)^{-\alpha(t')-1}$ & $-q^{\alpha_s \alpha_t}\frac{(q;q)_\infty (q^{1-\alpha_s-\alpha_t};q)_\infty}{(q^{1-\alpha_s};q)_\infty (q^{1-\alpha_t};q)_\infty}$ \\ 
         \hline
         Fixed transferred\linebreak momenta asymptotic ($s\rightarrow\infty,\ t=\text{const.}$)& $g^2(-gs)^{\alpha(t)}\Big[G(t)+\frac{\ln(-gs)}{s}G_1(t)$ $+\frac{1}{s}G_2(t)\Big]$&$-(-as)^{\alpha_t}\left[H(t)- \frac{1}{s}H_2(t)\right]$\\
         \hline
         Fixed angle asymptotic ($s,t\rightarrow\infty,\ s/t=\text{const.}$)& $e^{-\gamma\ln(2g)\ln(st)/2}$ & $-e^{\ln(-as) \ln(-at)/\ln q}(q;q)_\infty$\\
         \hline
         Resonances& $g^{k+1}\sum_{l=0}^k[-s\alpha'(s)]^l\frac{C_{k-l}(t)}{[k-\alpha(s)]^{l+1}}$ & $-\frac{q^{k\alpha_t}}{(\alpha_s-k)\ln q} \frac{(q^{1-k-\alpha_t};q)_k}{(q^{1-k};q)_{k-1}}$\\
         \hline
    \end{tabular}\label{tab1}
    }
\end{table}
\renewcommand{\arraystretch}{1}

Using this table, we can compare DAMA's and Dual-log's properties.

For ($s\rightarrow\infty,\ t=\text{const.}$), both amplitudes satisfy Regge asymptotic behavior; however, they differ at the next to leading order.

For ($s,t\rightarrow\infty,\ s/t=\text{const.}$), the Dual-log model maintains Regge-pole behavior while DAMA scales (following quark counting laws). This smooth transition from a ``soft'' to a ``hard'' regime is an important DAMA feature and may give insight into the confinement problem.

Both amplitudes show the linear spectrum of resonances ($\alpha_{s,t} = k , k \in \mathbb{N} $), but the properties of the poles are different. While the Dual-log model has simple poles, DAMA's poles are of order $k+1$. Moreover, it follows from DAMA that the number of resonances is finite - they melt beyond a certain mass. In Dual-log models, termination of resonances is associated with an ``ionization point''. Note that string theories predict an infinite number of narrow resonances. Any link between strings and a realistic dual model is of interest to theory.

We argue that q-analogs (quantum deformations and q-deformed algebra) may shed more light on the nature of these classes of dual models.

\appendix

\section{Fixed angle asympotic of DAMA}\label{app1}
Here we will derive a fixed angle asymptotic ($s,t\rightarrow\infty,\ s/t=\text{const.}$) of DAMA (for more details, see Refs.~\refcite{jenk79,jenk80,fiore1999fixed}). Regge amplitude has form \eqref{DAMA}:
\begin{equation*} 
A(s,t)=\int_0^1
dx\left(\frac{x}{g}\right)^{-\alpha(s')-1}\left(\frac{1-x}{g}\right)^{-\alpha(t')-1}.
\end{equation*}
We will consider $\alpha(s')$, $\alpha(t')$ in such forms:
\begin{equation*}
\alpha(s')=\alpha_0-\frac{\gamma}{2}\ln(-s(1-x))=\alpha_0-\frac{\gamma}{2}\ln(-s)-\frac{\gamma}{2}\ln(1-x),
\end{equation*}
\begin{equation*}
\alpha(t')=\alpha_0-\frac{\gamma}{2}\ln(-tx)=\alpha_0-\frac{\gamma}{2}\ln(-t)-\frac{\gamma}{2}\ln(x).
\end{equation*}
Hence
\begin{align*} 
A(s,t)&=\int_0^1
dx\left(\frac{x}{g}\right)^{-\alpha_0+\gamma\ln(-s)/2+\gamma\ln(1-x)/2-1}\left(\frac{1-x}{g}\right)^{-\alpha_0+\gamma\ln(-t)/2+\gamma\ln(x)/2-1}=\\
&=g^{\alpha_0-\gamma\ln(-s)/2+1}g^{\alpha_0-\gamma\ln(-t)/2+1}\int_0^1
dx \cdot x^{-\alpha_0+\gamma\ln(-s)/2-1} g^{-\gamma\ln(1-x)/2} \cdot \\
&\qquad\cdot(1-x)^{-\alpha_0+\gamma\ln(-t)/2-1}g^{-\gamma\ln(x)/2}e^{\gamma \ln(x) \ln (1-x)}=\\
&=g^{2\alpha_0-\gamma\ln(st)/2+2}\int_0^1
dx \cdot x^{-\alpha_0+\gamma\ln(-s)/2-\gamma\ln(g)/2-1} \cdot \\
&\qquad\cdot(1-x)^{-\alpha_0+\gamma\ln(-t)/2-\gamma\ln(g)/2-1} e^{\gamma \ln(x) \ln (1-x)}\\
\end{align*}
In the domain $x\in[0,1]$ we have $\exp{\left(\gamma \ln(x) \ln (1-x)\right)}\simeq 1$. Hence
\begin{align*}
A(s,t)&\simeq g^{2\alpha_0-\gamma\ln(st)/2+2}\int_0^1
dx \cdot x^{-\alpha_0+\gamma\ln(-s)/2-\gamma\ln(g)/2-1} \cdot \\
&\qquad\cdot(1-x)^{-\alpha_0+\gamma\ln(-t)/2-\gamma\ln(g)/2-1}=\\
&=g^{2\alpha_0-\gamma\ln(st)/2+2} B(-\alpha_0+\frac{\gamma}{2}\ln(-s)-\frac{\gamma}{2}\ln(g),-\alpha_0+\frac{\gamma}{2}\ln(-t)-\frac{\gamma}{2}\ln(g)).
\end{align*}
Using Stirling's approximation
\begin{equation*}
B(y,z)\simeq \sqrt{2\pi}\frac{y^{y-1/2}z^{z-1/2}}{(y+z)^{y+z-1/2}}=\sqrt{\frac{2\pi(y+z)}{yz}}\frac{1}{(1+z/y)^y(1+y/z)^z}.
\end{equation*}
For $y=-\alpha_0+\frac{\gamma}{2}\ln(-s)-\frac{\gamma}{2}\ln(g)$, $z=-\alpha_0+\frac{\gamma}{2}\ln(-t)-\frac{\gamma}{2}\ln(g)$, when $s,t\rightarrow\infty,\ s/t=\text{const.}$ 
\begin{equation*}
\sqrt{\frac{2\pi(y+z)}{yz}}\frac{1}{(1+z/y)^y(1+y/z)^z}\simeq\sqrt{\frac{2\pi(y+z)}{yz}}\frac{1}{2^{y+z}}.
\end{equation*}
Therefore,
\begin{align*}
A(s,t)&\simeq g^{2\alpha_0-\gamma\ln(st)/2+2} \sqrt{\frac{2\pi\left(\frac{\gamma}{2}\ln(-s)+\frac{\gamma}{2}\ln(-t)\right)}{\frac{\gamma}{2}\ln(-s)\frac{\gamma}{2}\ln(-t)}}2^{2\alpha_0-\gamma\ln(st)/2+\gamma\ln(g)}\sim\\
&\sim g^{-\gamma\ln(st)/2} \frac{1}{\sqrt{\ln(s)}}2^{-\gamma\ln(st)/2} \sim \frac{1}{\sqrt{\ln(s)}} (st)^{-\gamma\ln(g)/2}(st)^{-\gamma\ln(2)/2}\sim\\
&\sim (st)^{-\gamma\ln(2g)/2}.
\end{align*}
This result coincides with an equation \eqref{Dscale}.

\section*{Acknowledgements}
I thank L. Jenkovszky for useful discussions.

\bibliographystyle{ws-mpla}
\bibliography{references}

\end{document}